\documentstyle[aps,multicol,prl,epsfig]{revtex}
\def\FigW{7cm}
\def\FigH{4.3cm}

\newcommand{\beq}{\begin{equation}}
\newcommand{\eeq}{\end{equation}}

\begin{document}

\begin{title}
{\bf Families of Matter-Waves  for Two-Component Bose-Einstein Condensates}
\end{title}

\author{P.G. Kevrekidis$^{1}$, H. Nistazakis$^2$, D.J. Frantzeskakis$^{2}$, Boris A.
Malomed$^3$ and R.\ Carretero-Gonz\'{a}lez$^4$}
\address{$^{1}$ Department of Mathematics and Statistics,University of Massachusetts, Amherst MA 01003-4515, USA \\
$^{2}$ Department of Physics, University of Athens,
Panepistimiopolis,Zografos, Athens 15784, Greece \\
$^{3}$ Department of Interdisciplinary Studies, Faculty of Engineering, Tel
Aviv University, Tel Aviv 69978, Israel \\
$^4$ Nonlinear Dynamical Systems Group, San Diego State University, 
San Diego, CA 92182-7720}
%\title{}
\maketitle

\begin{abstract}
We produce several families of solutions for two-component nonlinear 
Schr\"{o}dinger/Gross-Pitaevskii equations. 
These include domain walls and
the first example of an antidark or gray soliton in the one component, bound
to a bright or dark soliton in the other. Most of these solutions are
linearly stable in their entire domain of existence. Some of them are
relevant to nonlinear optics, and all to Bose-Einstein condensates (BECs).
In the latter context, we demonstrate robustness of the structures in the
presence of parabolic and periodic potentials (corresponding, respectively,
to the magnetic trap and optical lattices in BECs).
\end{abstract}
%
%\PACS{{03.75,-b}{Matter waves} \and 
%      {52.35.Mw}{Nonlinear phenomena: waves, wave propagation, 
%                and other interactions}
%  }
%}
%\titlerunning{Matter-waves for two-component BECs}
%\authorrunning{P.G.\ Kevrekidis {\it et al.}}
 %end of abstract
%
%\maketitle

%\begin{multicols}{2}

\vspace{2mm}

The recent progress in experimental and theoretical studies of Bose-Einstein
condensates (BECs) \cite{review} has made matter-wave solitons physically
relevant objects. One dimensional (1D) dark \cite{dark} and bright 
\cite{hulet} solitons have been observed in experiments, and possibilities for
the observation of their multidimensional counterparts were predicted 
\cite{BaizOstr}. Further study of matter-wave solitons is a subject of profound
interest, not only from a theoretical perspective, but also for 
applications, as there are possibilities to coherently manipulate such
robust structures in matter-wave devices, e.g., atom chips, 
which are analogs of
the existing optical ones \cite{Folman}. On the other hand, many results
obtained for optical solitons as fundamental nonlinear excitations in
optical fibers and waveguides %, which were studied in vast literature 
(see, e.g., recent reviews \cite{buryak,kivpr}) suggest the possibility to
search for similar effects in BECs.

A class of physically important generalizations of the nonlinear 
Schr\"{o}dinger (NLS) equation for optical media, or its BEC counterpart, the
Gross-Pitaevskii equation (GP), is based on their multi-component versions.
In particular, 
%as concerns coupled GP equations describing multi-component BECs, 
the theoretical work has already gone into studying ground-state solutions 
\cite{shenoy,esry} and small-amplitude excitations \cite{excit} of the order
parameters in multi-component BECs. Additionally, the structure of binary
BECs \cite{Marek}, including the formation of domain walls in the case of
immiscible species, has also been studied \cite{Marek,healt}; 1D bound
dark-dark \cite{obsantos} and dark-bright \cite{anglin} soliton complexes,
as well as spatially periodic states \cite{decon}, were predicted too.
Experimental results have been reported for mixtures of different spin
states of $^{87}$Rb \cite{myatt} and mixed condensates \cite{dsh}. Efforts
were made to create two-component BECs with different atomic species, such
as $^{41}$K--$^{87}$Rb \cite{KRb} and $^{7}$Li--$^{133}$Cs \cite{LiCs}.

In this work, we report novel solitons in the context of coupled
two-component GP equations. These solutions correspond to new families of
solitons even for the NLS equations per se, hence they are also interesting
as nonlinear waves in their own right. We start by demonstrating their
existence in the context of two coupled NLS equations. Some of them are
relevant as new solitons in nonlinear-optical models as well. We also
demonstrate that all the solutions proposed herein persist in the presence
of the magnetic trap and optical lattice (OL), i.e., parabolic and
sinusoidal potentials \cite{tromb}, which are important ingredients of
experimental BEC setups.

Assuming that the nonlinear interactions are weak relative to the
confinement in the transverse dimensions, the transverse size of the
condensate is much smaller than its lengths. In this case, the BEC is a
\textquotedblleft cigar-shaped\textquotedblright\ one, and the GP equations
take an effectively 1D form \cite{GPE1d}: 
\begin{equation}
i\frac{\partial u_{j}}{\partial t}=-\frac{m_{1}}{m_{j}}\frac{\partial
^{2}u_{j}}{\partial x^{2}}
+\sum_{k=1}^{2}a_{jk}|u_{k}|^{2}u_{j}+V_{j}(x)u_{j},\qquad (j=1,2)
\label{teq1}
\end{equation}
where $u_{j}(x,t)$ are the mean-field wave functions of the two species, $t$
and $x$ are, respectively, measured in units of $2/\omega _{1\perp }$ and
the transverse harmonic-oscillator length $a_{1\perp }\equiv \sqrt{\hbar
/(m_{1}\omega _{1})}$ ($m_{j}$ and $\omega _{j\perp }$ are the mass and
transverse confining frequency of each species), while $\hbar \omega
_{1\perp }/2$ is the energy unit. The coefficients $a_{jk}$ in Eq.\ (\ref%
{teq1}), related to the three scattering lengths $\alpha _{jk}$ (note that 
$\alpha _{12}=\alpha _{21}$) through $a_{jk}=4\pi m_{1}(\alpha
_{jk}/a_{1\perp })(m_{j}+m_{k})/(m_{j}m_{k})$, account for collisions
between the atoms belonging to the same ($a_{jj}$) and different ($a_{jk}$, 
$j\neq k$) species; they are counterparts of the, respectively, self-phase
and cross-phase modulation in nonlinear optics. While in optics only
specific ratios of the nonlinear coefficients are relevant (such as 
$a_{ij}/a_{ii}=2$ or $a_{ij}/a_{ii}=2/3$ \cite{buryak}), in the BEC context
the interactions are tunable \cite{esry,decon}, especially because they can
be modified by means of the Feshbach resonance (i.e., by magnetic field
affecting the sign and magnitude of the scattering length of the interatomic
collisions) \cite{inouye}. The Feshbach resonance allows one to switch 
between attractive and repulsive interaction \cite{feshb}, and even to
switch it periodically in time, by means of an ac magnetic field, which
allows to create a self-confined 2D BEC without the magnetic trap 
\cite{Fatkh}.

In this work, we consider the case with $m_{1}=m_{2}\equiv m$ and 
$a_{11}=a_{22}$, which corresponds to the most experimentally feasible
mixture of two different hyperfine states of the same atom species, or,
approximately, to different isotopes of the same alkali metal, trapped in
the potential including the magnetic trapping and OL components: 
\begin{equation}
V_{1}(x)=V_{2}(x)\equiv V(x)=\left( \Omega ^{2}/2\right) \,x^{2}+V_{0}\sin
^{2}(kx+\phi ).  \label{teq2}
\end{equation}
In Eq.\ (\ref{teq2}), $\Omega ^{2}\equiv 2\omega _{x}^{2}/\omega _{\perp
}^{2}$ ($\omega _{1x}=\omega _{2x}\equiv \omega _{x}$ are the confining
frequencies in the axial direction) and $V_{0}$ (measured in units of the
recoil energy \cite{tromb}) set the respective potential strengths, $k$ is
the wavenumber of the interference pattern of the laser beams forming the
OL, and $\phi $ is an adjustable phase parameter ($\phi \in \{0,\pi /2\}$).
To estimate physical parameters, we resort to a mixture of two different
spin states of $^{87}$Rb, confined in a trap with the transverse frequency 
$\omega _{1\perp }=183$ rad/s, which implies that the length and time units
are $2\mu $m and $5.46$ ms, respectively. 
% (implying, e.g., that the above mentioned time step 
%is $27.3\mu$s in physical units).

We consider a rather general case, in which 
the inter-atomic interactions in the
first species are repulsive [therefore, we will use the normalization 
$a_{11}\equiv +1$ in Eqs. (\ref{teq1})], 
%corresponding, e.g., to different spin
%states of $^{87}$Rb or a $^{23}$Na, in which the scattering lengths 
%are in the proportion $\alpha_{11}:\alpha_{12}:\alpha_{22}=1.03:1:0.97$ 
%or $\alpha_{11}:\alpha_{12}:\alpha_{22}=1:1.035:1.035$ respectively
while in the other species they may be either attractive or repulsive. As
concerns the interactions between the different species, they are,
typically, repulsive. Nevertheless, in the case of two different spin states
of the same atom species, the Feshbach resonance between such states is
possible too (experimental studies of the Feshbach resonance in this case
are currently in progress \cite{private}), therefore attractive
inter-species interactions may be relevant, and this case is 
also considered
below. The solutions reported herein, and their
existence and stability regimes are summarized in Table 1.

%: \newline
%
%{\bf (i)} Domain walls for repulsive interactions in both species and
%repulsive interactions between the two species. \newline
%
%{\bf (ii)} Bound dark-antidark solitons in the same case 
%(which corresponds to the experiment reported in Ref. \cite{myatt}). \newline
%
%{\bf (iii)} Bound dark-gray solitons for repulsive inter-species interactions and attractive
%interactions between the two species. \newline
%
%{\bf (iv)} Bound bright-antidark solitons for repulsive interactions in the one
%component and attractive interactions in the other, while interactions between the two species  %are attractive. \newline
%
%{\bf (v)} Bound bright-gray solitons for the same case, but with repulsive interactions 
%between the two species. 

%\begin{tabular}{|l|l|l|l|} \hline
%\centering
%{\em Types of solitons} & $a_{11}$ & $a_{22}$ & $a_{12}$ \\ \hline\hline
%Domain walls & $+$ & $+$ & $+$ \\ \hline
%Dark-antidark & $+$ & $+$ & $+$ \\ \hline
%Dark-gray & $+$ & $+$ & $-$ \\ \hline
%Bright-antidark & $+$ & $-$ & $-$ \\ \hline
%Bright-gray & $+$ & $-$ & $+$ \\ \hline
%\end{tabular}
%\\
%\\

\vspace{2mm}

\centerline{
\begin{tabular}{|l||l|l||l|l|}
%\begin{center}
\hline
%\begin{center}
\centering
\emph{Types of solitons} & \multicolumn{2}{c||}{\emph{Existence}}
& \multicolumn{2}{c|}{\emph{Stability}} \\ \hline
& $a_{22}$ & $a_{12}$ & $a_{22}$ & ~~~$a_{12}$ \\ \hline
Domain wall & $+$ & $+$ & $+1$ & $>1$ \\ \hline
Dark-antidark & $+$ & $+$ & $+1$ & $(0,0.7]$ \\ \hline
Dark-gray & $+$ & $-$ & $+1$ & $[-0.83,0)$ \\ \hline
Bright-antidark & $-$ & $-$ & $-1$ & $(-1,0)$ \\ \hline
Bright-gray & $-$ & $+$ & $-1$ & $>0$ \\ \hline
%\end{center}
\end{tabular}
\newline
\newline
}
{Table 1: Existence and stability of structures in the binary BEC.
In the \textquotedblleft existence" column, }${+/-}${\ indicates the
repulsive/attractive character of the respective inter-atomic interaction
which is necessary for the solution to exist. The \textquotedblleft
stability" column indicates the sign of the coefficient }$a_{22}$ (we
normalize $a_{11}\equiv +1$, and set $a_{22}=\pm 1$) and an interval of the
values of $a_{12}$ for which the solution is stable.

\vspace{5mm}

In most cases the existence and stability of the solution families is
investigated numerically. The numerical method was implemented as follows:
we first seek stationary solutions by means of Newton iterations 
%(with the finite-difference step $\Delta x=0.3$ and free boundary conditions), 
which are applied to the steady-state equations
%\[
$\mu u_{j}=-u_{j,xx}+\sum_{k=1}^{2}a_{jk}|u_{k}|^{2}u_{j}+V(x)u_{j}$ ($\mu $
is the chemical potential). %\]
Subsequently, we perform the linear-stability analysis of the obtained
soliton solutions $u_{j}^{(0)}(x)$, setting the perturbed solution to be
%\begin{equation}
$u_{j}=e^{-i\mu t}\left[ u_{j}^{(0)}(x)+\epsilon \left( b_{j}e^{-i\omega
t}+c_{j}e^{i\omega ^{\ast }t}\right) \right] $,\label{teq5} %\end{equation}
where $\omega \equiv \omega _{r}+i\omega _{i}$ is a (generally, complex)
perturbation eigenfrequency. Then, the ensuing linear stability problem 
\cite{sulem} is solved for the eigenfrequencies and eigenfunctions 
$\{b_{j},c_{j}^{\ast }\}$. %from Eq.\ (\ref{teq5}) 
Whenever the solution is unstable, we also examine its evolution in direct
simulations of the full equations (\ref{teq1}), using a fourth-order
Runge-Kutta time integrator with the time step $dt=0.005$
($27.3\mu$s in physical units). To initiate the
instability development, a uniformly distributed random perturbation of 
amplitude $\sim 10^{-4}$ was typically added to an unstable solution. 

We now examine in detail
 the solutions shown in Table 1. First, in the absence of the
external potential, a family of domain walls can be found in an exact
form for the special case, $a_{12}=3a_{11}=3a_{22}$: 
\begin{equation}
u_{j}(x,t)=Ae^{-i\mu t}\left[ 1+(-1)^{j}\tanh (\eta x)\right] ,  \label{teq3}
\end{equation}
where the chemical potential is $\mu =4a_{12}A^{2}$, and $\eta
^{2}=2a_{12}A^{2}$ (they follow the pattern of domain-wall solutions found
long ago in the context of coupled Ginzburg-Landau equations \cite{GL}).
These solutions exist only if $a_{12}>0$ and $\mu >0$. Similar patterns were
found in Ref. \cite{healt}
%no analytical form was presented in that work. 
and other related structures were also predicted to occur in higher
dimensions \cite{dwnew}. 
%In Fig.\ \ref{tfig1-2} (left panel), we display this solution, 

We have confirmed the existence and stability of the domain walls
 by direct numerical
simulations (not shown here), using numerical continuation to extend them to
the case $a_{12}\neq 3a_{11}$, 
%(recall we set $a_{11}\equiv 1$), 
where the
analytical solution is not available. We have thus found that the domain
walls exist
and are stable for values of $a_{12}$ down to $a_{12}=1$. 
%as is shown in the left panel of Fig.\ \ref{tfig1-2}.
% (where the normalization is $\mu=4$). 
The case $a_{12}=2$ is relevant to nonlinear optics; stability of the 
domain wall
family for this case was suggested by recent numerical results obtained
for a similar discrete coupled-NLS model \cite{DWus}. Here, we find that
these solutions are robust as well for other values of $a_{12}$, and, as
will be shown below, also in the presence of the external potential in Eq.
(\ref{teq2}).

%\begin{figure}[tbp]
%\twoimages[width=\FigW,height=\FigH]{twoc3c.ps}{twoc4c.ps}
\begin{figure}[tbp]
\includegraphics[width=\FigW,height=\FigH]{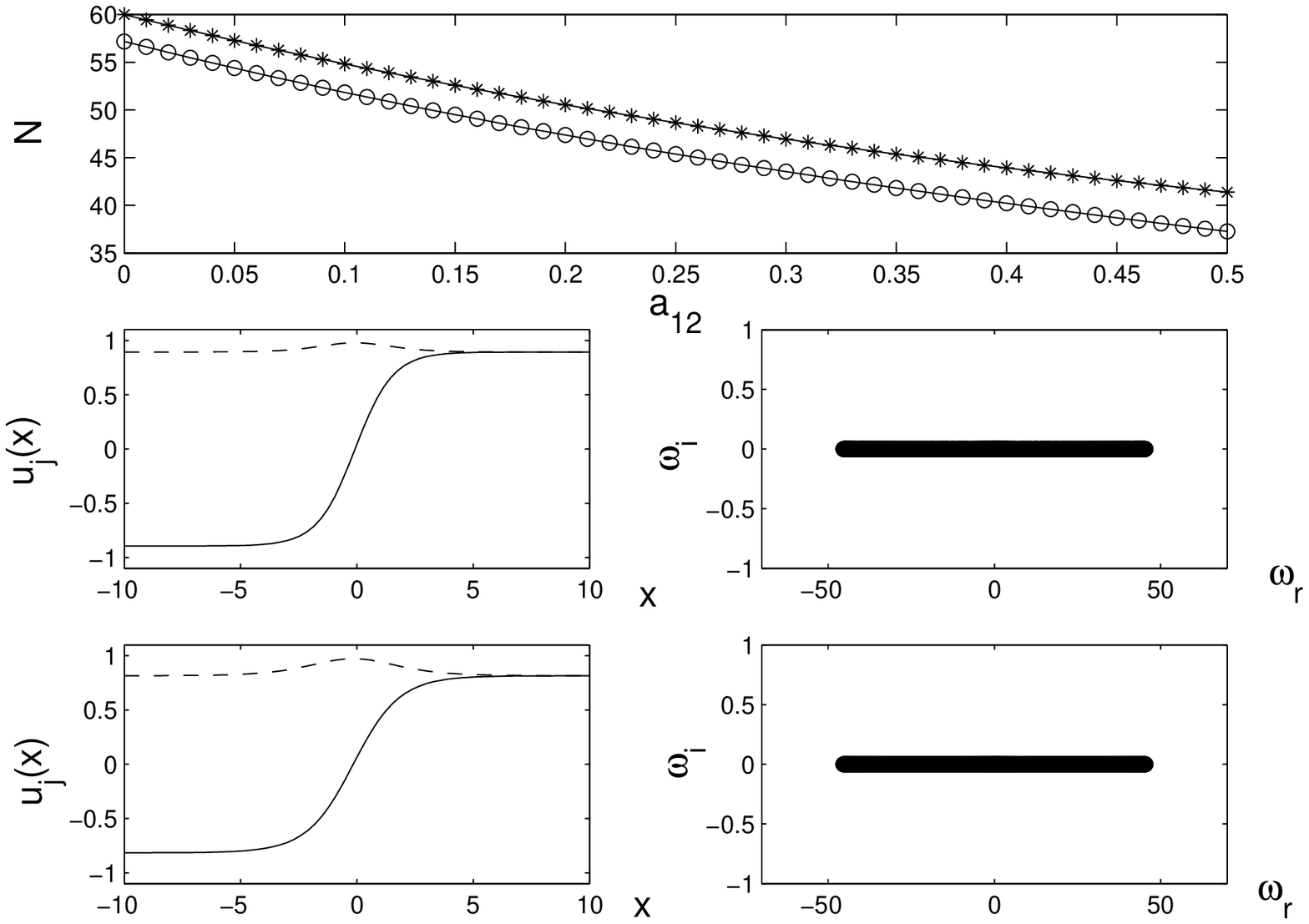}
\includegraphics[width=\FigW,height=\FigH]{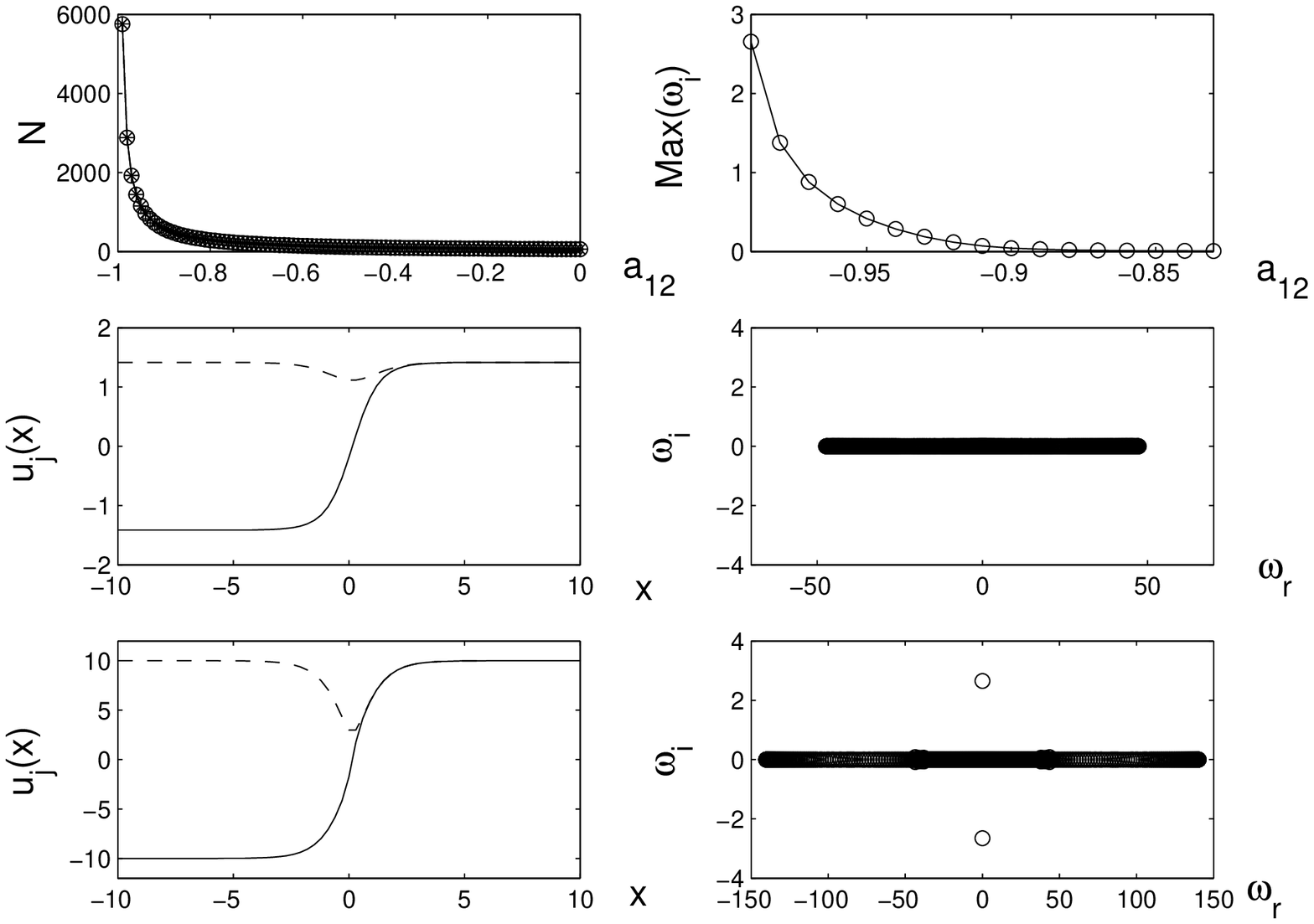}
\caption{ Left: The top panel shows the number of atoms (in normalized
units) in each component (circles and stars) of the dark-antidark soliton. 
The middle and bottom left panels show the spatial profiles of
this solution for $a_{12}=0.2$ and $a_{12}=0.5$, while the right panels show
the respective spectral planes $(\protect\omega _{r},\protect\omega _{i})$
of the corresponding eigenfrequencies, as found from the linearized 
equations ($\omega _{i}= 0$ implies linear stability). 
Right: The top panels show the number
of atoms in each component of the dark-gray soliton
and the instability growth versus $a_{12}$. The middle and bottom
left panels show the spatial profiles of this solution for $a_{12}=-0.5$ and 
$a_{12}=-0.99$, while the right panels show the respective spectral planes. }
\label{tfig1-2}
\end{figure}

Proceeding to the other solutions in Table 1, we first note that, upon
fixing $a_{11}\equiv +1$, we set $a_{22}=\pm 1$ for the repulsive and
attractive interactions in the second species. Although setting $\left\vert
a_{22}\right\vert $ to be $1$ formally limits the generality of the results,
it has been checked that taking $\left\vert a_{22}\right\vert \neq 1$ 
produces similar results to those displayed here. 

Given that we have assumed the first component as being always self-repulsive, 
we look for
solutions starting 
from the uncoupled limit ($a_{12}=0$) by taking an initially
uniform distribution in this component, $u_{1}\equiv 1$. If 
the second component
is also repulsive, we take a dark soliton as its initial configuration. If
it is self-attractive, a bright soliton is initially set in it.

The case of the self-repulsion and dark soliton in the second component,
with the inter-species coefficient being \emph{repulsive} too, $a_{12}>0$
gives rise to a stationary \textit{antidark soliton} (i.e., a hump on
top of a nonzero flat background) in the first component, see the left panel
in Fig.\ \ref{tfig1-2}. It is easy to understand this structure, as the
atoms in the first component, being repelled by the matter in the second
one, concentrate in an effective potential well generated by the dark
soliton (void) in it. Antidark solitons 
are well-known to occur when higher-order effects
(such as third-order dispersion) or a saturable nonlinearity are present in
the single-component NLS equations, which is possible in optics \cite{kivpr}.
In that case, the antidark soliton is 
usually described by a KdV-type asymptotic
equation (for the elevation on top of the flat background), and is not
stationary, running at the respective velocity of sound \cite{kivpr}. The
two-species soliton with the antidark component, presented here, is the first
example of a stationary antidark soliton, 
that we are aware of, in a model without
higher-order nonlinearities and dispersions. It is also the \emph{first
prediction} of antidark solitons in BECs, which suggests 
that an experimental verification would be of particular interest.

If, on the contrary, the interaction between the two self-repelling species
is \emph{attractive} ($a_{12}<0$; right panel in Fig.\ \ref{tfig1-2}), then
the void (dark soliton) in the second component effectively repels the
matter in the first one, and thus generates a dip, i.e.\ a \textit{gray
soliton} in it (for a detailed description, see Ref.\ \cite{kivpr}). Such
solitons exist in the regular NLS equation, but there (as well as in other
instances of their presence that we are aware of) they travel at a nonzero
speed (the faster the shallower the dip is), while here the gray solitons
are stationary.

%\begin{figure}
%\twoimages[width=\FigW,height=\FigH]{twoc2c.ps}{twoc9c.ps}
%\caption{
%Left:
%Branch of the bright-gray coupled solitons.
%Right:
%The continuation of the bright-antidark branch versus $V_{0}$, for $k=1$, 
%$\phi =0$ and $a_{12}=-0.5$. The cases of $V_{0}=0.5$ and $V_{0}=1$ are shown
%in the middle and bottom panels.} 
%\label{tfig5-6} 
%\end{figure}

The bound dark-antidark two-component states persist for $0<a_{12}\leq 0.7$,
while the dark-gray bound-state branch continues down to $a_{12}=-1$,
getting appreciably unstable in the region $-1<a_{12}<-0.83$. In the latter
case, the time evolution (not shown here) leads to formation of \emph{moving}
dark-gray soliton complexes. 

\begin{figure}[tbp]
%\twoimages[width=\FigW,height=\FigH]{twoc5c.ps}{twoc2c.ps}
\includegraphics[width=\FigW,height=\FigH]{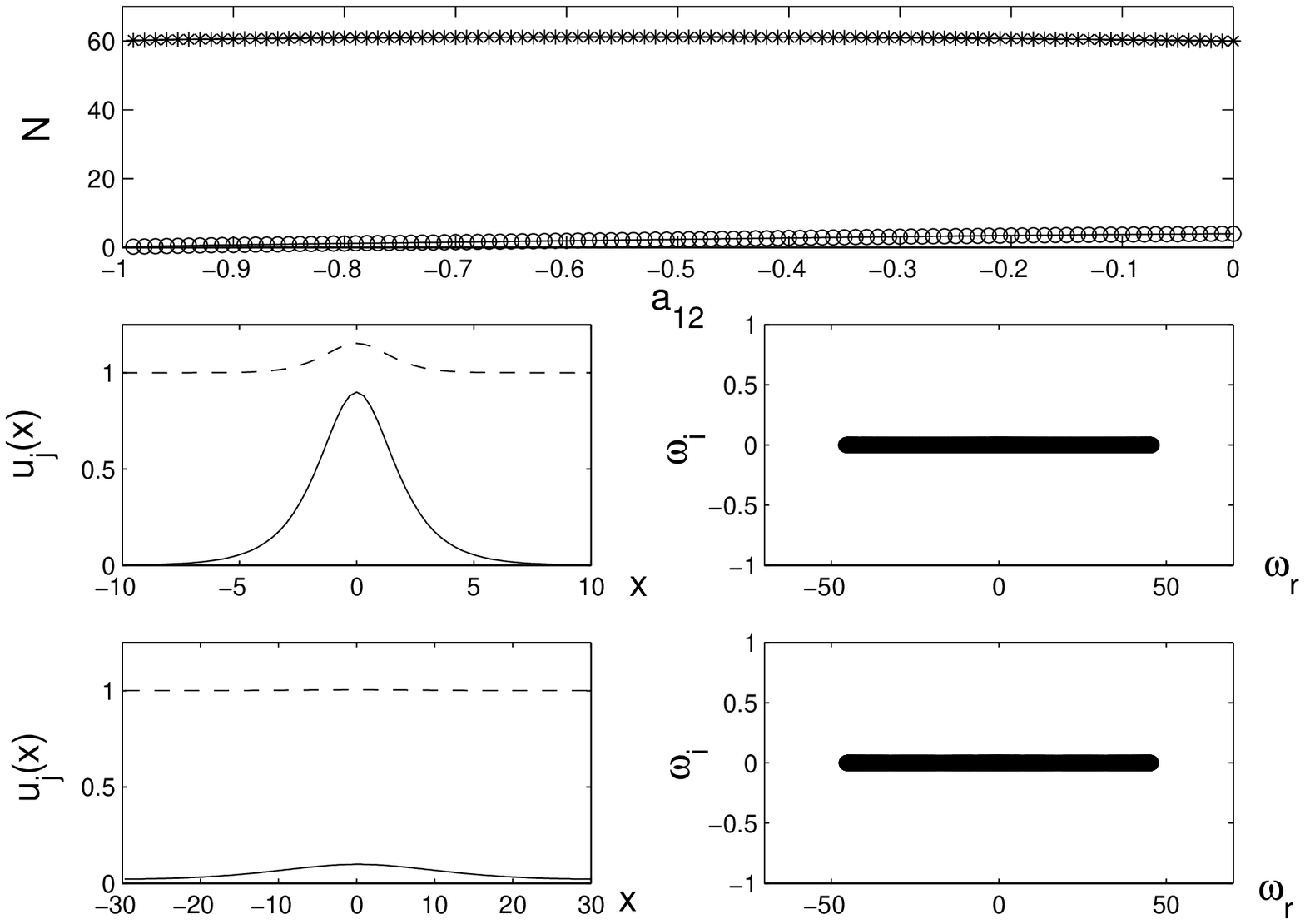}
\includegraphics[width=\FigW,height=\FigH]{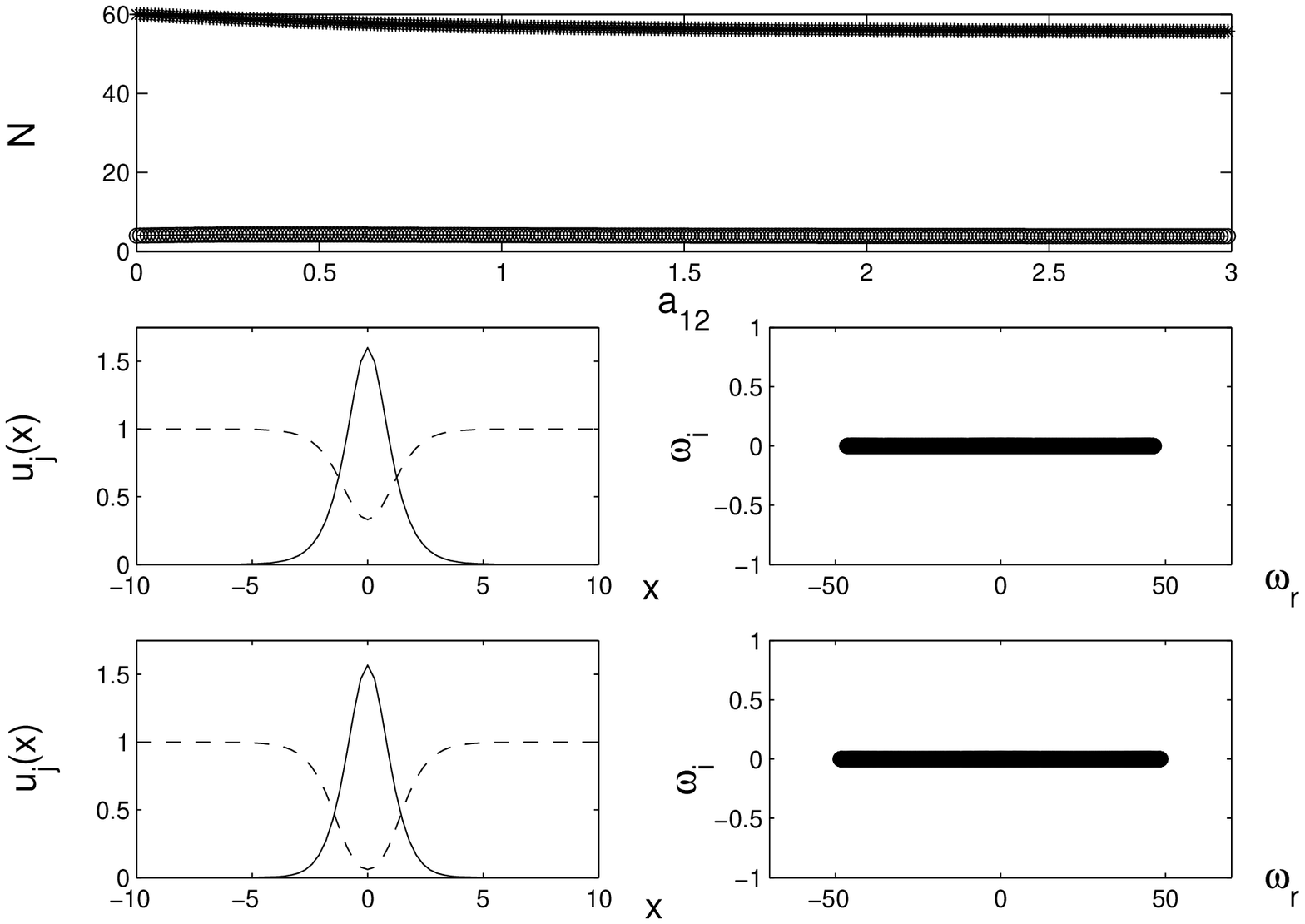}
\caption{ Left: Same as Fig.\ \protect\ref{tfig1-2} but for the
bright-antidark solitons. The middle and bottom panels are for $a_{12}=-0.5$
and $a_{12}=-0.99$, respectively. Right: The branch of coupled bright-gray
solitons. The middle and bottom panels are for $a_{12}=0.2$ and $0.5$,
respectively. }
\label{tfig3-4}
\end{figure}

Similar results were obtained for the last two branches of the new
solutions: if the second component is now self-attractive, then the
attractive interaction between the BEC species ($a_{22}=-1,a_{12}<0$; see
left panel in Fig.\ \ref{tfig3-4}) gives rise to \textit{bright-antidark
solitons} (whose existence is explained by the fact that the bright soliton
attracts matter in the other component). Such a type of two-component
solitons was predicted in optics \cite{djf}, but there it cannot exist
without the third-order dispersion. In our case, this solution branch
terminates at $a_{12}=-1$, since both states become completely flat. 

Lastly, if the bright soliton repels matter in the other component, it
naturally induces a dip in it, thus generating a \textit{bright-gray soliton}
in the case of $a_{22}=-1,a_{12}>0$ (right panel Fig.\ \ref{tfig3-4}). It is
noteworthy that this bright-gray branch is extremely robust; we were able to
follow such stable solutions down to $a_{12}=-3$.

\begin{figure}[tbp]
\includegraphics[width=\FigW,height=\FigH]{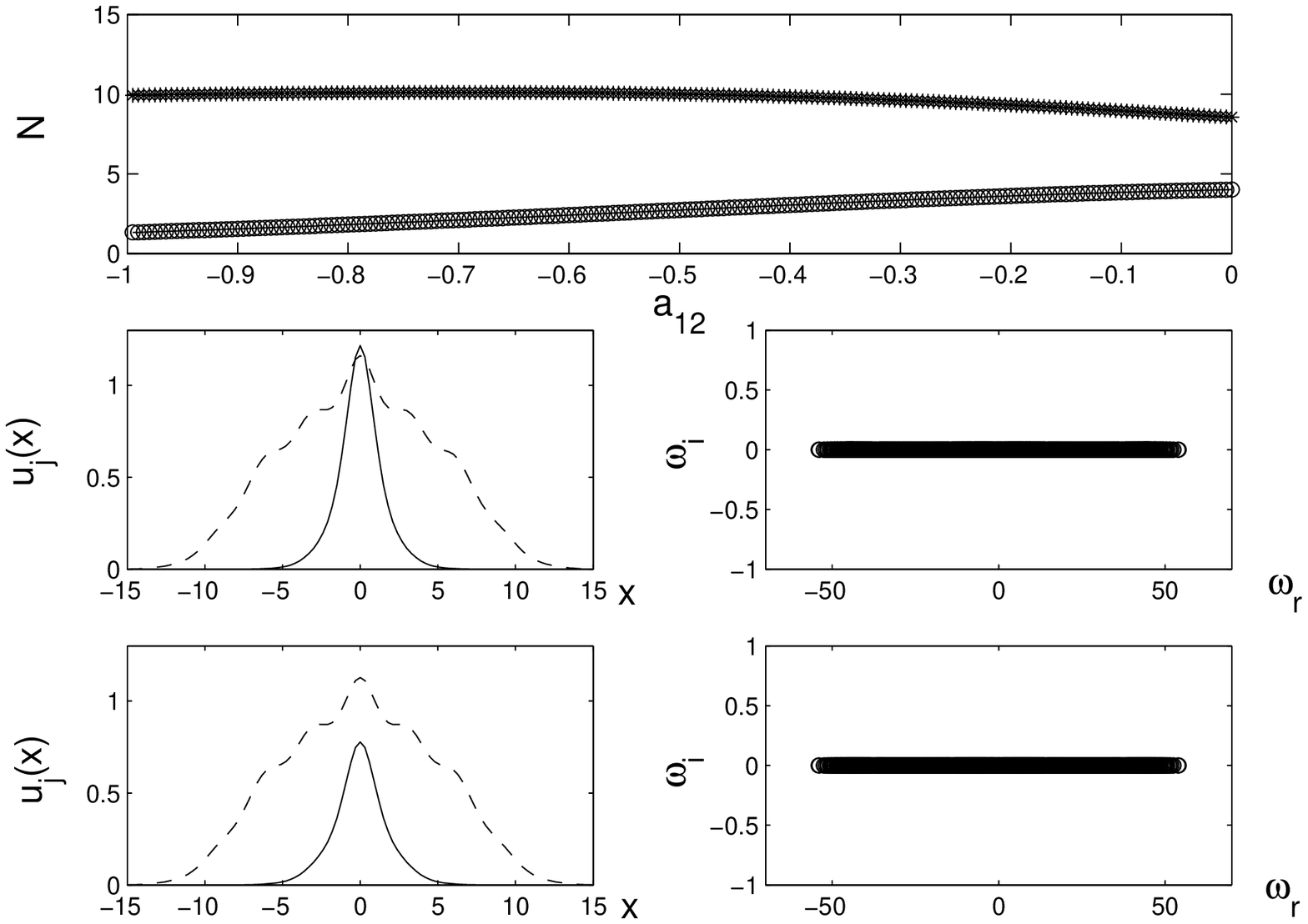}
\includegraphics[width=\FigW,height=\FigH]{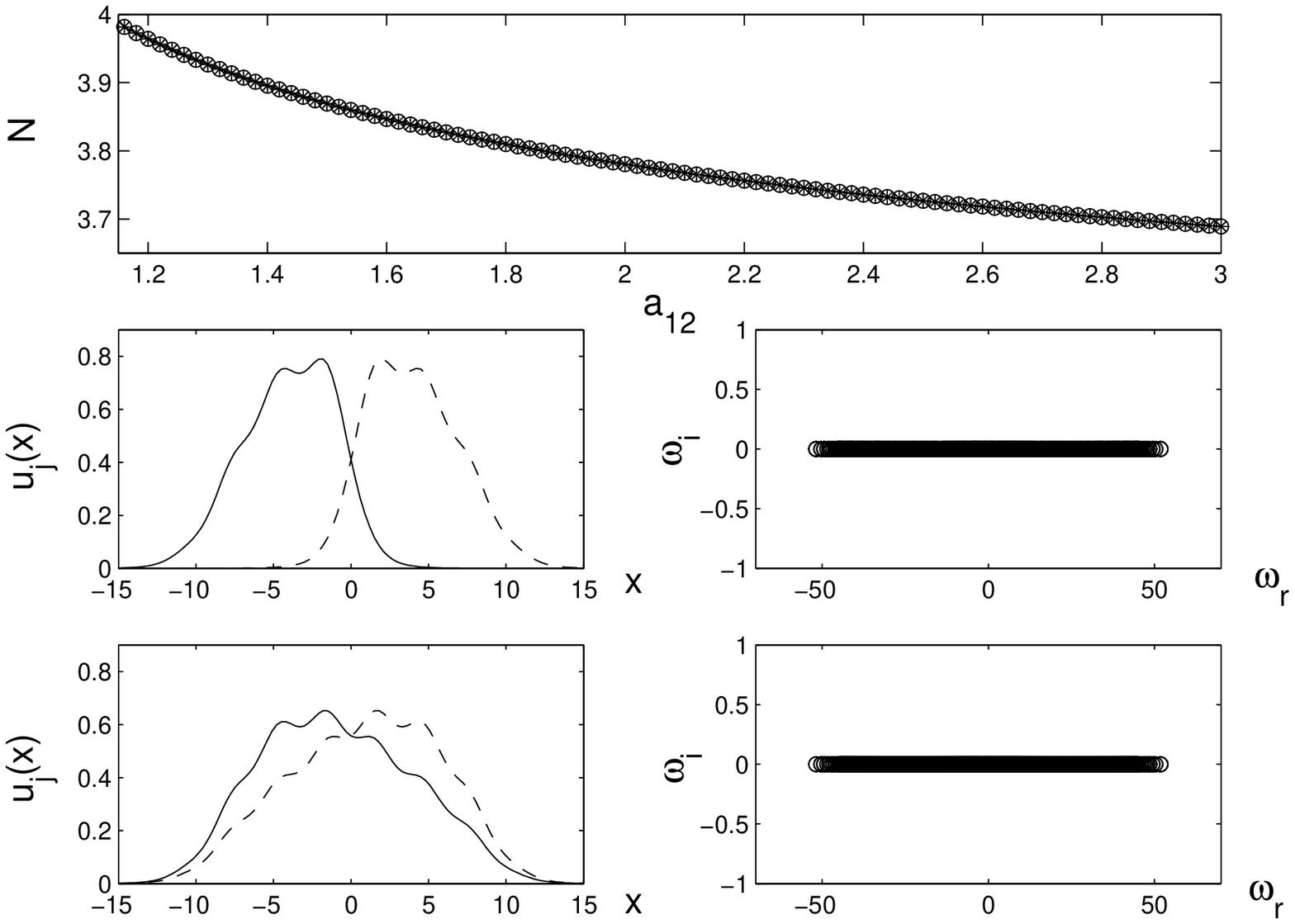}
%\twoimages[width=\FigW,height=\FigH]{twoc13c.ps}{twoc14c.ps}
\caption{ Left: The bright-antidark branch (the middle and bottom panel are for 
$a_{12}=-0.5$ and $-1$, respectively) in the presence of the potential 
$V(x)=0.01x^{2}+0.5\sin ^{2}(x)$, see Eq.\ (\protect\ref{teq2}). Right: 
The same
for the domain-wall solutions (the middle and bottom panels are for 
$a_{12}=3 $ and $2$, respectively). }
\label{tfig7}
\end{figure}

We have also examined these novel solutions in the presence of the OL [i.e.,
for $V_{0}\neq 0$, $\Omega =0$ in Eq.\ (\ref{teq2})]. We have found them to
be typically stable when $\phi =0$, and unstable when $\phi =\pi /2$ (as
might be expected, since the solution is then placed, respectively, at a
minimum and maximum of the potential). %A typical example concerns the
%bright-antidark branch [with $k=1$ in Eq.\ (\ref{teq2})] 
%is shown in the right panel in Fig.\ \ref{tfig5-6}. 
%where the continuation is performed over the optical lattice strength $V_{0}$, for $a_{12}=-
%0.5$. 
However, there are exceptions to this rule. 
%in the presence of the optical lattice). 
For instance, dark-antidark bound solitons for $a_{12}=0.5$ are unstable at 
$\phi =0$, in intervals $0.04\leq V_{0}\leq 0.055$ and $0.09\leq V_{0}\leq
0.29$. This oscillatory instability, involving a quartet
of eigenvalues, will be examined in detail elsewhere.

We have also identified all the solution branches in the presence of the
magnetic trap, as well as in the case when both the magnetic
trap and the OL are present.
The branches are extremely robust in the presence of the magnetic trap. In
particular, Fig.\ \ref{tfig7} shows the branches for the bright-antidark
bound state (left panel) and domain walls 
(right panel), which are \textit{always
stable} in the presence of the combined potential with $\Omega ^{2}=0.02$
and $V_{0}=0.5$. In particular, for a mixture of two different spin states
of $^{87}$Rb, the confining frequencies corresponding to this value of 
$\Omega $ are $\omega _{x}=18.3$ rad/s and $\omega _{1\perp }=183$ rad/s in
the transverse and axial directions, respectively. Then, the four cases
shown in Fig.\ \ref{tfig7} correspond to the mixture containing, in the
first and second species, $2\times 10^{4}$ and $6\times 10^{3}$ atoms 
($a_{12}=-0.5$), $2\times 10^{4}$ and $3\times 10^{3}$ atoms ($a_{12}=-1$), 
$7.4\times 10^{3}$ in each component ($a_{12}=3$), and, finally, $7.6\times
10^{3}$ atoms in each component ($a_{12}=2$), respectively.

\begin{figure}[tbp]
%\twoimages[width=\FigW,height=\FigH]{twoc11c.ps}{twoc11instc.ps}
\includegraphics[width=\FigW,height=\FigH]{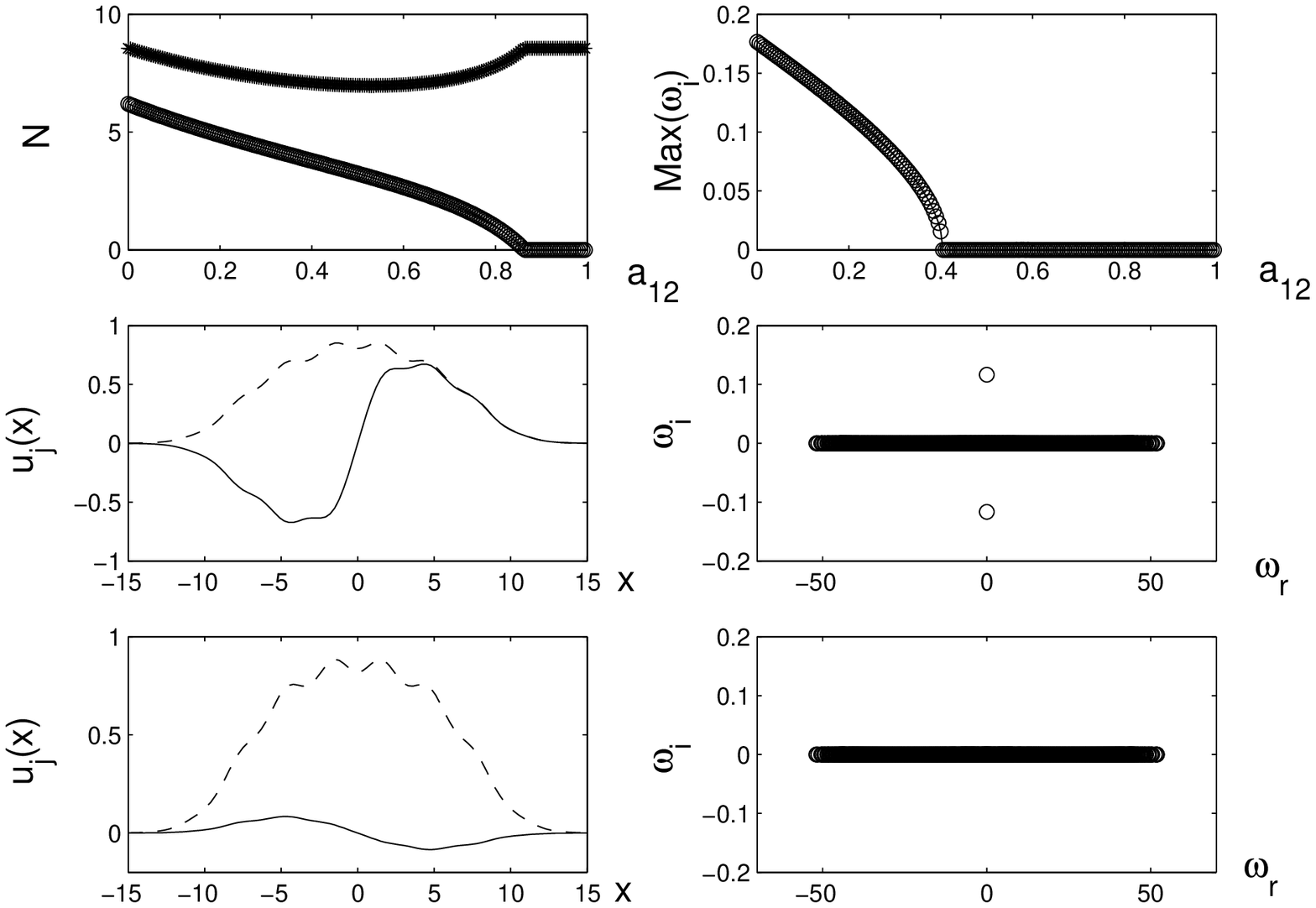}
\includegraphics[width=\FigW,height=\FigH]{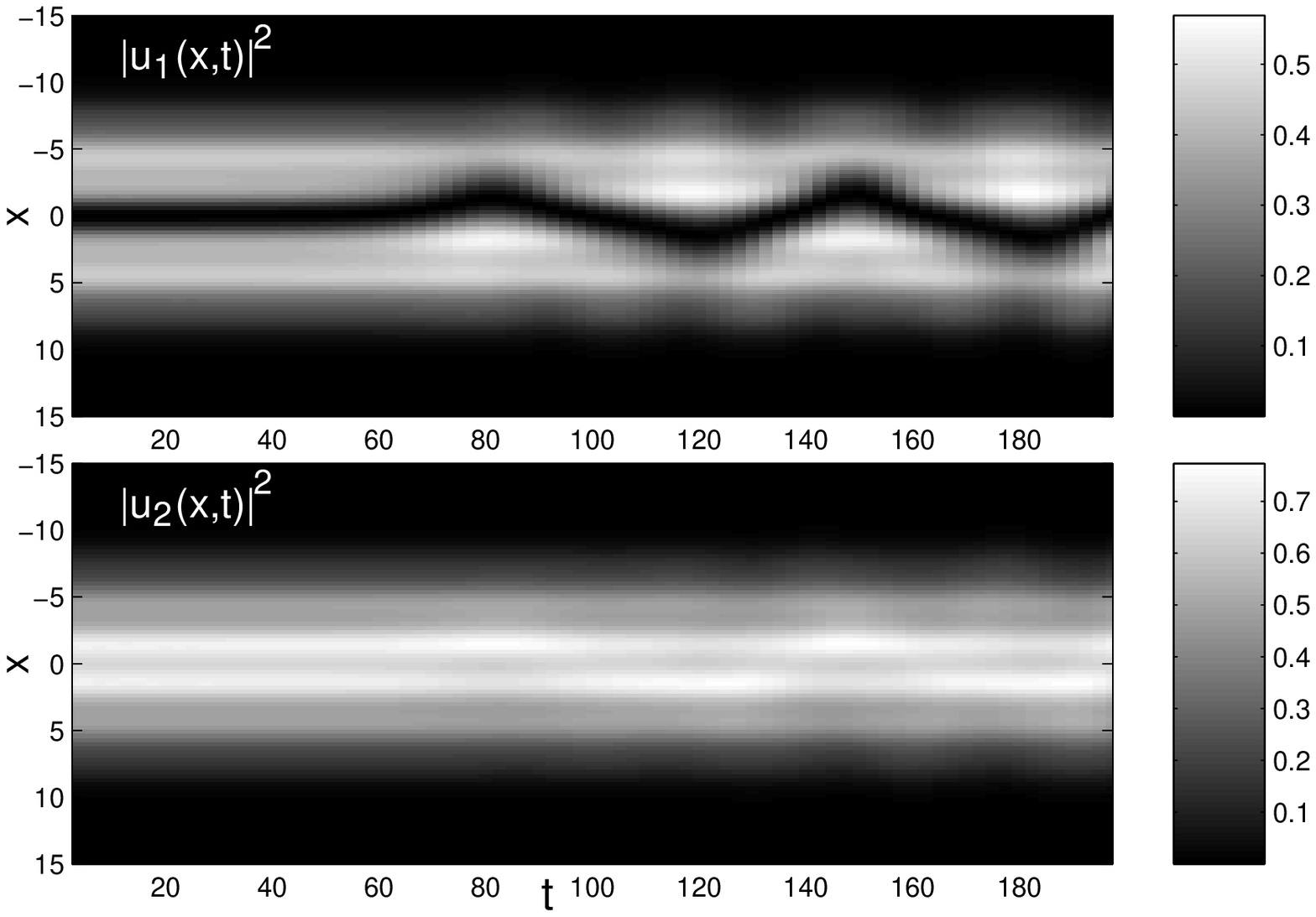}
\caption{ Left: The dark-antidark solution branch for $\phi =
\pi /2$. The branch is unstable at $a_{12}<0.4$ (an example
is displayed for $a_{12}=0.2$, in the middle panel), and
then it gets stabilized (see an example for $a_{12}=0.86$
in the bottom panel). Right: Time evolution of
the unstable soliton in the case $a_{12}=0.2$ with a random perturbation of
an amplitude $\sim 10^{-4}$ added to the initial condition. 
The instability leads to
an oscillating dark soliton. The time unit is $5.46$ ms, hence the onset
of instability occurs at $\approx 0.33$ s. }
\label{tfig8}
\end{figure}

Finally, we notice that the increasing interaction between the components
can play a stabilizing role for solutions that are unstable in the OL with 
$\phi =\pi /2$ (Fig.\ \ref{tfig8}). The dark-antidark coupled state serves as
an example, being unstable for $0<a_{12}<0.4$ and stable for $a_{12}>0.4$
(left panel). The instability evolution is shown in the right panel for 
$a_{12}=0.2$: the dark soliton becomes mobile and starts to oscillate in the
combined potential. For the physical example mentioned above, the
numbers of atoms in the two components corresponding to the latter case
are $10^{4}$ and $1.6\times 10^{4}$, respectively.

In conclusion, we have presented a number of novel families of composite
solutions in the generic two-species BEC model. In particular, the first
possibility to create an antidark soliton in BECs is predicted. In most
cases, the compound solitons and domain walls are very robust, keeping the
stability in the presence of the parabolic and periodic potentials. In some
cases, the new solutions are relevant to optical models as well. 
It would be of
interest to look for such states experimentally.

\vspace{5mm}

%\acknowledgments 
{\bf Acknowledgments}: We appreciate a valuable discussion with D.S. Hall. This
work was supported by a UMass FRG, NSF-DMS-0204585 and the Eppley Foundation
(PGK), the Special Research Account of the University of Athens (HEN, DJF),
the Binational (US-Israel) Science Foundation under grant No. 1999459 (BAM)
and the San Diego State University Foundation (RCG).

%\end{multicols}

\end{document}